\documentclass[twocolumn,superscriptaddress,floatfix,prl,aps,showpacs]{revtex4}
\usepackage{graphicx}
\begin{document}
\bibliographystyle{apsrev}
\def\sa{\section}
\def\sb{\subsection}
\def\sc{\subsubsection}

\def\ind{\ \ \ \ }

\def\be{\begin{equation}}
\def\ee{\end{equation}}

\def\bea{\begin{eqnarray}}
\def\eea{\end{eqnarray}}

\def\ba{\begin{array}}
\def\ea{\end{array}}
 
\def\nn{\nonumber}

\def\ben{\begin{enumerate}}
\def\een{\end{enumerate}}

\def\fn{\footnote}

\def\rd{\partial}
\def\rot{\nabla\times}

\def\r{\right}
\def\l{\left}

\def\gt{\rightarrow}
\def\cf{\leftarrow}
\def\bw{\leftrightarrow}

\def\ra{\rangle}
\def\la{\langle} 
\def\bla{\big\langle}
\def\bra{\big\rangle}
\def\Bla{\Big\langle}
\def\Bra{\Big\rangle}

\def\ddt{{d\over dt}}

\def\rdt{{\rd\over\rd t}}
\def\rdx{{\rd\over\rd x}}

\def\bb{\begin{thebibliography}{99}}
\def\eb{\end{thebibliography}}
\def\bit{\bibitem}

\def\bc{\begin{center}}
\def\ec{\end{center}}

\title{ Artificial electric field  in Fermi Liquids }
\author{Ryuichi Shindou}
\affiliation{Department of Physics, University of California, 
Santa Barbara, California 93106, USA}
\affiliation{Department of Physics, University of Tokyo,
Bunkyo-ku, Tokyo 113-8656, Japan.}
\author{Leon Balents}
\affiliation{Department of Physics, University of California, 
Santa Barbara, California 93106, USA}
\date{\today} 

\begin{abstract}
  Based on the Keldysh formalism, we derive an effective Boltzmann
  equation for a quasi-particle associated with a particular Fermi
  surface in an interacting Fermi liquid.  This provides a many-body
  derivation of 
  Berry curvatures in electron dynamics
  with spin-orbit coupling, which has received much attention in recent
  years in non-interacting models.  As is well-known, the Berry
  curvature in momentum space modifies na\"ive band dynamics via an
  ``artificial magnetic field'' in momentum space.  Our Fermi liquid
  formulation completes the reinvention of modified band dynamics by
  introducing in addition an {\it artificial electric field}, related to
  Berry curvature in frequency and momentum space.  We show explicitly
  how the artificial electric field affects the renormalization factor
  and transverse conductivity of interacting $U(1)$ Fermi liquids with
  non-degenerate bands.  Accordingly, we also propose a method
  of {\it momentum resolved Berry's curvature detection} in terms of angle
  resolved photoemission spectroscopy (ARPES). 
\end{abstract}



\maketitle


Recent experimental developments in spintronics, notably several
measurements of evidently {\sl intrinsic} anomalous Hall effect in
ferromagnetic metals~\cite{tokura}, and observations of the 
spin Hall effect in semiconductors~\cite{asc,jung}, 
have highlighted the importance of intrinsic spin orbit
effects in conductors.  Theoretically, there has been a renewal of
interest in the fundamental dynamics of Bloch electrons in periodic
solids taking into account spin orbit interactions, following ideas
pioneered by Karplus and Luttinger~\cite{KL}, 
and elegantly formulated by Sundaram
and Niu~\cite{sundaram}. In particular, in a non-interacting band
picture, the low energy Hilbert space is restricted to states in the
conduction band containing the Fermi surface {\sl at each quasimomentum
  $k$ along this surface}.  This {\sl local constraint}, like other
local constraints in physics~\cite{dimer}, 
generates a gauge symmetry in momentum 
space, corresponding to the freedom to redefine Bloch functions by a
momentum-dependent phase factor (or $SU(2)$ matrix in the case of
two-fold degenerate bands).  Consequently, a wavepacket analysis of
non-interacting Bloch states shows that this structure induces {\it 
artificial magnetic field (a.m.f.)}
, which 
contributes a {\it Lorentz force in $k$-space}, mathematically analogous
to the action of the real magnetic field ${\bf b}$ in real
$R$-space~\cite{sundaram}.  The a.m.f.  ${\cal B}_{\alpha}$    
and its associated magnetic gauge field ${\cal A}_{\alpha}$ 
are defined from the periodic part of the Bloch 
wavefunction of conduction bands $|u_{\alpha}\ra$: 
${\cal B}_{\alpha,i} \equiv i\epsilon_{ijm}\partial_{k_j}
{\cal A}_{\alpha,m}$ and ${\cal A}_{\alpha,i} \equiv \bla u_{\alpha} 
\big|\partial_{k_{i}} u_{\alpha}\bra$.  
The fundamental framework of Fermi liquid theory suggests that these
conclusions persist in realistic metals, in which the Coulomb
interactions between electrons are not weak.  Haldane has argued that
this is indeed the case for quasiparticles at the Fermi surface, with
$|u_{\alpha}\ra$ replaced by {\it the eigenvector of the inverse of 
the Hermitian part of the retarded (advanced) Green function}~\cite{FDMH}.
In this letter, we identify an {\sl additional} manifestation of this
gauge structure unique to {\sl interacting} Fermi liquids.  In 
particular, because the renormalized eigenstates thus introduced are in 
general dependent on frequency $\omega$, it is natural to introduce an {\sl
  artificial electric field (a.e.f.)} and {\sl electrostatic field} as  
${\cal E}_{\alpha, i} \equiv i\big(\partial_{\omega}{\cal
  A}_{\alpha,i}-\partial_{k_i} {\cal A}_{\alpha,0}\big)$ and 
${\cal A}_{\alpha,0} \equiv \bla u_{\alpha} \big|\partial_{\omega}
u_{\alpha}\bra$ respectively.  
These quantities vanish in the absence of interactions, since the
$\omega$-dependence of the quasiparticle states arises from that
of the (collisional) self-energy.

In the following, we present explicit calculations that demonstrate this 
a.e.f. indeed has physical effects.  In addition, our
Keldysh formulation provides a many-body derivation of an effective 
kinetic equation embodying the 
{\it entire} emergent gauge structure, and a systematic
procedure for actually carrying out the projection into the low energy
band.  We concentrate on metallic materials lacking either inversion or
time-reversal symmetry, in which case the bands are non-degenerate and
the gauge structure is $U(1)$.  For such a $U(1)$ Fermi liquid (FL), we 
demonstrate that both a.m.f. and a.e.f.,   
which are estimated ``on-shell'' (see Eq.~(\ref{aem}) in combination with 
Eqs.~(\ref{curv},\ref{newbasis}) for their more precise definitions),  
in fact enter into the effective equation of motion (EOM) for a 
quasi-particle;  
\begin{eqnarray} \frac{dR}{dT} &=& {\bf v} + 
\big({\cal B}_{\alpha} -  
{\cal E}_{\alpha}\times {\bf v}\big) \times \frac{dk}{dT}, \nonumber \\
\frac{dk}{dT} &=& - {\bf e} + {\bf b} \times \frac{dR}{dT}.  
\label{EOMquasi} 
\end{eqnarray} 
${\cal E}_{\alpha}$ thus provides {\it another source of the 
Lorentz force in $k$-space}, in addition to the
contribution of ${\mathcal B}_{\alpha}$ field identified in
Ref.~\onlinecite{sundaram}.  It therefore modifies the transverse
current carried by a quasiparticle, and hence the intrinsic anomalous
Hall conductivity of a conducting ferromagnet to
\begin{equation}
  \label{eq:ahe}
\sigma_{jk} =  \epsilon_{jkl}\cdot 
\frac{e^2}{\hbar}\sum_{\alpha} \int  
\frac{dk}{(2\pi)^d}\ ({\cal B}_{\alpha} - {\cal E}_{\alpha}
\times{\bf v}_{\alpha})_{l}n_{\alpha},
\end{equation}
where the sum $\alpha$ is over bands, and $n_{\alpha}$ 
is the distribution function.  In particular, 
Eq.~(\ref{eq:ahe}) can be also proven to be consistent with 
the many-body formulation of the transverse conductivity \cite{IM,FDMH}, 
more directly by using Ward identites which relate the current 
vertex correction to single quasi-particle Green functions 
\cite{fut}. 
As another physical consequence of these 
artificial electromagnetic fields (a.e/m.f), 
we will also show that the {\it linear response} 
of the momentum-resolved single-particle density of states   
to applied 
electromagnetic fields is  
characterized by these artificial 
fields. 
Specifically, in the presence of ${\bf e}$ and ${\bf b}$, 
the {\it quasiparticle renormalization factor} of a $U(1)$ FL acquires 
the following topological corrections:
\begin{eqnarray}
Z' = 1-\frac{1}{2}\left({\cal B}_{\alpha}\cdot{\bf b} 
+ {\cal E}_{\alpha}\cdot{\bf e}\right). \label{z'}
\end{eqnarray} 
This in principle makes it possible to detect the {\sl distribution} of
these ``artificial'' fields in the momentum space, using ARPES
experiments.

We now turn to the derivation of the above results and the
gauge-invariant effective kinetic equation.  We begin with a quite general
semi-microscopic model in which the electronic spectrum is described by
a $k\cdot p$ type expansion about some (arbitrary) point in the
Brillouin zone.  One may keep as many bands as are deemed close enough
in energy to be relevant to the physics, and our arguments do not depend
upon the order in the expansion in $k$.  Familiar examples would be the
multi-band Luttinger models for commonly studied semiconductors, in
which the natural expansion is about the $\Gamma$ point.  In principle
one should choose the expansion point so that there are no low-energy
band crossings intervening between it and the Fermi surface, and there
could be difficulties in treating open Fermi surfaces with the
periodic structure of the quasimomentum.  However, while a fully
microscopic treatment would be desirable, our approach is nevertheless
quite general and moreover we speculate that our final results naturally
extend and apply to arbitrary $U(1)$ FLs.

The advantage of this formulation is that we can Fourier transform in
the usual way to a continuous real space coordinate $r$, i.e. $k
\rightarrow -i\nabla_r$.  The non-interacting  Hamiltonian is thereby 
expressed in terms of the fermionic ``envelope fields''
$\psi_{\alpha}(r)$:
$  \label{eq:ham0}
  {\cal H}_{0}=\sum_{\alpha,\alpha'}  
  \int dr\psi^{\dagger}_{\alpha}(r)
  [\hat{H}_{0}(-i\nabla_{r},r)]_{\alpha\alpha'}\psi_{\alpha'}(r)$,  
where the band index $\alpha^{(\prime)}$ runs 
from $1$ to  $N_b$.  
A specific information about lattice, orbital and spin-orbit 
couplings are encoded into the matrix structure of $[\hat{H}_{0}]$.  
We will employ a natural form of the interaction, ${\cal H}_{1}= 
\sum\int\int \psi^{\dagger}_{\alpha_1}(r_1) \psi^{\dagger}_{\alpha_2}(r_2)
V_{\alpha_1\alpha_2\alpha^{'}_2\alpha^{'}_1}(r_1,r_2)
 \psi_{\alpha^{'}_2}(r_2) \psi_{\alpha^{'}_1}(r_1)$, though the form of
 our results does not depend in detail upon this.

Bearing in mind this Hamiltonian, ${\cal H} = {\cal H}_{0} + {\cal
  H}_1$, we can proceed with the Keldysh formalism, beginning with an
analysis of the spectral function $[\hat{\sf A}(r,t|r',t')]_{(\alpha|\alpha')} = \bla 
\big\{\psi_{\alpha}(r,t),\psi^{\dagger}_{\alpha'}(r',t')\big\}_{+}\bra$.
It obeys~\cite{rammer,KB} 
\begin{eqnarray}
\label{eq:kdeom}
[\hat{G}_{0}^{-1}-\hat{\Sigma}^{\rm HF}-\hat{\sigma},\hat{\sf A}]_{\otimes,-}
\equiv[\hat{L},\hat{\sf A}]_{\otimes,-}= [\hat{\Gamma},{\rm 
  Re}\hat{G}]_{\otimes,-}  \rightarrow \hat{0}.   
\end{eqnarray}
Here the product $\otimes$ denotes the convolution w.r.t. time $t$,
space $r$ and band index $\alpha$, $(\hat{B}\otimes \hat{C})(1|1')
\equiv \int d\bar{1} \hat{B}(1|\bar{1})\cdot \hat{C}(\bar{1}|1')$.  The 
commutator here is defined by using this convolution:
$[\hat{B},\hat{C}]_{\otimes,-}\equiv \hat{B}\otimes \hat{C} - 
\hat{C}\otimes \hat{B}$. The bare Green function is as usual defined by
$\hat{G}^{-1}_{0}(1|1') \equiv 
[i\partial_{t_1}\hat{1} -
{\hat{H}_{0}(-i\nabla_{r_1},r_1)}]_{\alpha_1\alpha^{'}_1}\delta(t_1-t^{'}_1)\delta(r_1-r^{'}_1)$.
$\hat{\Sigma}^{\rm HF}(1|1')$ and $\hat{\sigma}(1|1')$ are the
Hartree-Fock part (temporally instantaneous) and the hermitian part of the
collisional self-energy (non-instantaneous), while $\hat{\Gamma}(1|1')$
is the anti-hermitian part of the collisional self-energy.
In a FL, the decay rate represented by the $[\hat{\Gamma},{\rm
  Re}\hat{G}]_{\otimes,-}$ in Eq.~(\ref{eq:kdeom}) is expected to be
$O((\omega-\mu)^2,T^2)$, due to the phase-space constraints inherent to
FL theory.  Because we focus on low temperature physics (and indeed
require the self-energy only to $O(\omega-\mu)$, the order that
determines the non-vanishing electric field at the Fermi surface 
\cite{comment2}), we
henceforth ignore this lifetime effect, replacing the right hand side in
Eq.~(\ref{eq:kdeom}) as indicated.

Even this {\sl dissipationless} Keldysh equation is still non-trivial
and involved, due to the presence of the {\it band index}, whose effect is
the central issue of this work.  In the following, we will argue a
general method of projecting out the irrelevant band indices and how
to obtain the reduced Keldysh (kinetic) equation only for the relevant bands.

The first step is to apply the
Wigner transformation, so that frequency and
momentum $(\omega,q)$ are introduced by the Fourier transformation for
the relative coordinate in space and time
$(t,r)=(t_1-t_{1'},r_1-r_{1'})$. Namely, $\hat{B}(\omega,q;T,R) = \int
drdt \hat{B}(r_1,t_1|r_{1'},t_{1'})e^{-iqr+i\omega t}$, where the center
of mass coordinate in space and time
$X=(T,R)\equiv(\frac{t_1+t_{1'}}{2},\frac{r_1+r_{1'}}{2})$ parameterize
phase space in combination with $Q=(\omega,q)$.
Then, following Ref.~\onlinecite{rammer}, the Keldysh equation in terms 
of $Q,X$ becomes
\begin{widetext}
\begin{eqnarray}
 -\ \big[\hat{L},\hat{\sf A}\big]_{-}
= \frac{i}{2}\ \big[\partial_{X_j}\hat{L},\partial_{Q_j}\hat{\sf A}\big]_{+} 
-\ \frac{1}{8}\ \Big(\big[\partial_{X_j}\partial_{X_k}\hat{L}
,\partial_{Q_j}\partial_{Q_k}\hat{\sf A}\big]_{-} 
-\ \big\{ X_k \leftrightarrow Q_k\big\} \Big) 
\ -\ \big\{ X_j \leftrightarrow Q_j\big\}.  \label{2nd}
\end{eqnarray}
\end{widetext}
The $j$ and $k$-summation over $0$ to $d$ are implicit from now on ($d$
is the spatial dimension of our system).  The (anti)commutator here is
defined in the sense of the product {\it only with respect to band
  indices}.  The non-local correlation effect which was previously
encoded in the space-time convolution is now partially taken into
account via the so-called gradient expansion in
$\partial_{Q}\partial_{X} \equiv - 
\partial_{\omega}\partial_{T} + 
\partial_{q_j}\partial_{R_j}$.  The expansion is justified close to
equilibrium, since all quantities are independent of $X$ in equilibrium.
We carry out the gradient expansion to 2nd order, at which 
${\cal E}$, ${\cal B}$ appear in the kinetic 
equation (see, for example, Eq.~(\ref{eff4})).

The Wigner-transformed 
{\it Lagrangian} $\hat{L}$, when diagonalized,   
$\hat{L}_d=\hat{U}^{\dagger}\cdot\hat{L}\cdot\hat{U}$, 
defines a renormalized energy dispersion 
by its eigenvalue and renormalized Bloch functions by its 
eigenbasis. 
Namely, in this new basis, 
Eq.~(\ref{2nd}) {\it at equilibrium} can be satified by 
those spectral functions having no off-diagonal 
components. The diaognal elements of 
$\hat{A}\equiv \hat{U}^{\dagger}\cdot\hat{\sf A}\cdot \hat{U}$
are verified a posteriori  
to become a delta-function whose argument is the eigenvalue 
of $\hat{L}$   
\begin{eqnarray}
A_{\alpha}\equiv A_{\alpha\alpha} 
= \delta(L_{d,\alpha})=\delta(\omega - E_{\alpha}(\omega,q))
\label{equi},  
\end{eqnarray}
where $E_{\alpha}(\omega,q)$ is the $\alpha$-th eigenenergy of $\hat{H}_{0} +
\hat{\Sigma}^{\rm HF} + \hat{\sigma}$. 

Out of equilibrium, 
the off-diagonal element also 
acquires small but finite weight, which we will take into
account in a systematic way below.  In this new basis,  the simple
derivatives in Eq.~(\ref{2nd}) is replaced by covariant ones,
\begin{eqnarray}
  \hat{U}^{\dagger}\cdot\big(\partial_{Q_j}\hat{L}\big)\cdot\hat{U} 
  &=& 
  \partial_{Q_j}\hat{L}_d +  
  \hat{\cal A}_{Q_j}\cdot\hat{L}_d - \hat{L}_d\cdot\hat{\cal A}_{Q_j},
  \nonumber \\  
  \hat{U}^{\dagger}\cdot\big(\partial_{X_j}\hat{\rm A}\big)\cdot\hat{U} 
  &=& 
  \partial_{X_j}\hat{A} +  
  \hat{\cal A}_{X_j}\cdot\hat{A} - \hat{A}\cdot\hat{\cal A}_{X_j}, \nonumber  \\
  \hat{U}^{\dagger}\cdot\big[\hat{L},\hat{\sf A}\big]_{-}\cdot\hat{U} 
  &=&\big[\hat{L}_d,\hat{A}\big]_{-}, \   
  \hat{\cal A}_{Q_j}\equiv\hat{U}^{\dagger}\partial_{Q_j}\hat{U}, 
  \label{newbasis}
\end{eqnarray}
$N_b$ times $N_b$ matrices $\hat{\cal A}_{Q_j}$ and $\hat{\cal A}_{X_j}$
are the {\it renormalized gauge field}, 
through which we will define 
a.e/m. f. later 
(see Eqs.~(\ref{aem},\ref{curv})).
As the off-diagonal components of the spectral function are expected to
be small in this new basis, they may be eliminated in favor of the
diagonal components.  Specifically, this occurs because the Keldysh
equation for the $A_{\alpha\beta}$ ($\alpha\neq \beta$) 
contains the term $\Delta_{\alpha\beta}
A_{\alpha\beta}$, with the {\sl direct band gap} 
$\Delta_{\alpha\beta} = E_{\alpha}-E_{\beta}$ as the
sole non-vanishing term at $0^{\rm th}$ order in the gradient expansion --
forcing $A_{\alpha\beta}$ to vanish in equilibrium.  
Accordingly, $A_{\alpha\beta}$ is {\it of at most 1st order in the 
gradient expansion, and it can be solved for in terms of 
the diagonal elements $A_{\alpha}$ by iteration}. 
By contrast, the diagonal components of the Keldysh equation
appear all at first and higher order in the gradient expansion. 
Substituting this iterative solution for  $A_{\alpha\beta}$  
into the diagonal equations, 
we can obtain a set of $N_b$ equations
for the diagonal parts $A_{\alpha}=A_{\alpha\alpha}$ 
alone, to the desired ($2^{\rm nd}$)
order in gradients.

Finally, we focus on a specific band ``$\alpha$'' which contains a Fermi
surface.  In the low frequency region ($|\omega-E_{\alpha}|\simeq
|\omega-\mu|\ll \min_{\beta}|\Delta_{\alpha\beta}|$), the diagonal
elements of the spectral functions {\it for the other bands} are
expected to have negligible weight.  This is because in equilibrium, in
the non-interacting limit, $A_{k\ne i}$ is sharply peaked at an energy
separated from the chemical potential by the direct band gap. With
interactions, or out of equilibrium, there may be some incoherent weight
around $\omega\simeq\mu$ but it is expected to be extremely small.  Thus
the other bands can also be eliminated, so that we obtain {\sl
  independent} Keldysh equations for the diagonal spectral functions.

Following this prescription, we obtain the differential equation
${\cal L}^{(1)}(A_{\alpha}) + {\cal L}^{(2)}(A_{\alpha})=0$ 
with~\cite{fut} 
\begin{eqnarray}
{\cal L}^{(1)}(A_{\alpha})
&=& \big(\partial_{X_j}(L_{d,\alpha} - {\cal M}_{\alpha})\big)
\big(\partial_{Q_j}A_{\alpha}\big) - 
\big\{X_j \leftrightarrow Q_j\big\}, \nonumber \\
{\cal L}^{(2)}(A_{\alpha})&=& \frac{1}{4}\big(\partial_{X_j}L_{d,\alpha}\big)
\big(\partial_{Q_j}\Omega^{\alpha}_{X_kQ_k}\big)A_{\alpha} \nonumber \\
&&- \big(\partial_{X_k}L_{d,\alpha}\big)
\Omega^{\alpha}_{Q_kX_j}\big(\partial_{Q_j}A_{\alpha}\big)
- \big\{X_j \leftrightarrow Q_j\big\} \nonumber \\ 
&&- \big\{X_k \leftrightarrow Q_k\big\} + \big\{X_k,X_j 
\leftrightarrow Q_k,Q_j\big\}.
\label{eff}
\end{eqnarray} 
Here the Berry curvatures associated with the $\alpha$-th band, i.e.
$\Omega^{\alpha}_{Q_jX_k}$ and its 3 counter parts, are defined in terms
of the renormalized gauge fields introduced in Eq.~(\ref{newbasis}),
e.g.  \bea \Omega^{\alpha}_{Q_jX_k}\equiv -i\sum_{\beta\ne
  \alpha}[\hat{\cal A}_{Q_j}]_{\alpha\beta}[\hat{\cal
  A}_{X_k}]_{\beta\alpha} + {\rm c.c.}. \label{curv} \eea ${\cal
  M}_{\alpha}$ in Eq.~(\ref{eff}) denotes an energy correction of the
1st order gradient expansion: ${\cal M}_{\alpha} \equiv
\frac{i}{2}\sum_{j=0}^{d}\sum_{\beta\ne \alpha} [\hat{\cal
  A}_{X_j}]_{\alpha\beta} \cdot\big(L_{d,\alpha}-L_{d,\beta}\big)
\cdot[\hat{\cal A}_{Q_j}]_{\beta\alpha} + {\rm c.c.}$.  The Zeeman
coupling energy between orbital momentum and external magnetic field,
which was previously found in the non-interacting case~\cite{sundaram},
is encoded into the $\sum_{j=1}^{d}$ terms in this energy correction.
On the other hand, the {\it temporal} term, i.e. $j=0$, describes a
coupling of the {\it electric dipole moment} of quasi-particles to the
external electric field, which is unique to an interacting FL \cite{fut}.

We can now extract the low energy behavior of the spectral function.  
The renormalized energy dispersion of the $\alpha$-th 
band $\epsilon(q;X)$ is determined from
\begin{equation}
  \label{eq:rendisp}
  L_{d,\alpha}-{\cal M}_{\alpha} 
\equiv \epsilon - E_{\alpha}(\epsilon,q;X) - {\cal
    M}_{\alpha}(\epsilon,q;X) = 0. 
\end{equation}
Indeed, {\sl neglecting} ${\cal L}^{(2)}$, the form
$A_{\alpha}=\delta(L_{d,\alpha}-{\cal M}_{\alpha})$, 
satisfies the Keldysh equation to
the 2nd order in the gradient expansion.  Therefore, 
without ${\cal L}^{(2)}$, the spectral function is sharply peaked at
$\omega=\epsilon$, with a weight given by the conventional
``renormalization factor'' $Z=[\partial_{\omega}(L_{d,\alpha}-{\cal
  M}_{\alpha})]^{-1}_{|\omega=\epsilon}$.  
However, due to the presence of
${\cal L}^{(2)}$, which contains terms 
proportional to $A_{\alpha}$ rather than
its derivative, the solution up to the 2nd order of the gradient
expansion acquires an {\it additional renormalization factor}
$Z'=1-\frac{1}{2}(\Omega^{\alpha}_{X_kQ_k})_{|\omega=\epsilon}$, 
and
\begin{eqnarray}
A_{\alpha} = Z'\delta(L_{d,\alpha}-{\cal M}_{\alpha})
= ZZ' \delta(\omega-\epsilon). \label{spec} 
\end{eqnarray}    

To unmask the meaning of this new term $Z'$, let us specialize to the situation
where the disequilibrium is generated by a physical (i.e. external)
electromagnetic gauge field $({\rm a}_0,{\bf a})$, with corresponding
physical electromagnetic fields 
${\bf b}=\nabla \times {\bf a}$, ${\bf e}=
\nabla_{R}{\rm a}_{0} - \partial_T {\bf a}$.  In this case, 
energy  
$\epsilon({\bf q},X)=\epsilon_0({\bf q}+{\bf a}(X))+{\rm a}_0(X)$, 
where $\epsilon_0$ is the (renormalized) band energy in equilibrium.  
Then, introducing canonical momentum and frequency $k=q+{\bf a}(T,R)$
and $\omega' \equiv\omega - {\rm a}_{0}(R)$~\cite{LL}, 
Eq.~(\ref{spec}) becomes 
$A_{\alpha} = Z Z' \delta(\omega'-\epsilon_0(k))$. 
Correspondingly, the Berry curvature appearing in 
the additional renormalization factor $Z'$ 
are expressed in terms of the following 
artificial electromagnetic fields (a.e/m.f.) estimated at $\omega=\epsilon$, 
\bea
(\Omega^{\alpha}_{k_mk_n})_{|\omega=\epsilon} 
= \epsilon_{mnj}{\cal B}_{\alpha,j}\ ,\ \ (\Omega^{\alpha}_{\omega k_j})_{|\omega=\epsilon}
={\cal E}_{\alpha,j}. \label{aem}
\eea 
Namely, $Z'$ simply reduces to 
an inner product between real electromagnetic fields and 
these a.e/m.f., as    
in Eq.~(\ref{z'}).  
This result indicates a renormalization of the ``angle resolved density
of states'' when both physical and a.e/m.f. 
are present.  A similar modification (dependent upon ${\cal
  B}_{\alpha}\cdot{\bf b}$ only) was proposed in Ref.~\onlinecite{Xiao}.

This theoretical observation leads us to propose the {\it
  momentum-resolved detection of a.e/m.f.} by
ARPES, by observing the modification of the spectral weight by small
applied electromagnetic fields.  This is no doubt quite difficult in
practice, but demonstrates that 
the systematic measurement of the artificial
fields is possible in principle.  

We now turn from the spectral function to the quasiparticle dynamics, by
deriving the {\it effective Boltzmann equation} for quasi-particles in a
$U(1)$ FL.  We start with the observation that the dissipationless
Keldysh equation in Eq.~(\ref{2nd}) holds also for the lesser Green's
function, ${\sf g}^{<}(1,1')\equiv  i\bla
\psi^{\dagger}_{\alpha'}(r',t')\cdot\psi_{\alpha}(r,t)\bra$, under the
same assumptions~\cite{KB}. 
Applying the same unitary transformation as
above, i.e. $\hat{g}^{<}\equiv \hat{U}^{\dagger}\cdot\hat{\sf g}^{<}\cdot \hat{U}$,
one then obtains exactly the same reduced 
Keldysh equation for the 
$(\alpha,\alpha)$-component of $\hat{g}^{<}$ as in Eq.~(\ref{eff}).  
This lesser Green's function may be decomposed according to
$\label{eq:gfd}
  g^{<}_{\alpha}(Q;X)=i A_{\alpha}(Q;X)\cdot f_{\alpha}(Q;X)$, 
where $f_\alpha(Q;X)$ is a generalized Fermi 
distribution~\cite{KB}.  
Inserting this into the reduced Keldysh equation thus obtained, 
we can readily obtain the EOM for $f_{\alpha}$, 
\bea
&0&=ZZ'\delta(\omega-\epsilon)\times \Big\{\partial_{X_j}(L_{d,\alpha}-
{\cal M}_{\alpha})\partial_{Q_j}f_{\alpha} 
 \nonumber \\ 
&-&\partial_{X_k}\Omega^{{\alpha}}_{Q_kX_j}\partial_{Q_j}f_{\alpha}\  +\ \partial_{Q_k}
\Omega^{\alpha}_{X_kX_j}\partial_{Q_j}f_{\alpha} - (X_j \leftrightarrow Q_j)\Big\}. \nonumber 
\eea
Because of the sharp peak in $A_\alpha \simeq \delta(\omega - \epsilon)$, 
when integrating this EOM w.r.t. $\omega$,
we get an effective Boltzmann equation 
for {\it a quasi-particle occupation number in $q-R$ space};  
$n_{\alpha}(q:R,T) \equiv f_{\alpha}(q,\epsilon:R,T)$, 
\begin{eqnarray}
&&\big(1-\Omega^{\alpha}_{T\epsilon} 
+ (\partial_{R_j}\epsilon)\ \Omega^{\alpha}_{q_j\epsilon} 
- (\partial_{q_j}\epsilon)\ \Omega^{\alpha}_{R_j\epsilon}\big)\ 
\partial_T n_\alpha \nonumber \\
&&=\ \big(\partial_{R_j}\epsilon 
+ (\partial_{T}\epsilon)\  \Omega^{\alpha}_{\epsilon R_j} 
- (\partial_{R_k}\epsilon)\ \Omega^{\alpha}_{q_kR_j} + \ \Omega^{\alpha}_{TR_j}  
 \nonumber \\
&&\hspace{0.5cm}\ +\  
(\partial_{q_k}\epsilon)\ \Omega^{\alpha}_{R_kR_j}\big)
\ \partial_{q_j}n_\alpha - \{ q_j \leftrightarrow R_j\} \label{eff3}. 
\end{eqnarray}      
Note that, although all the curvature terms in the above are now ``on-shell'' quantities, 
the partial  $q$, $R$ and $T$-derivatives  encoded there  
applies only on their {\it explicit} dependence  
and do {\it not} apply on their arguments of $\epsilon$.  

In the presence of external electromagnetic fields, 
this effective Boltzmann equation can be substantially  
simplified in terms of canonical momentum introduced above, 
\begin{eqnarray}
&& 0 = \partial_{T}n_{\alpha} \ 
+\ \Big[- {\bf e} + {\bf b}\times {\bf v} \nonumber \\
&&\ \ +\ {\bf b}\times({\bf e}\times {\cal B}^{\prime}_{\alpha}) 
- {\bf b}\times\big(({\bf b}\times{\bf v})\times{\cal B}^{\prime}_{\alpha})\Big]
\cdot \partial_{k}n_\alpha \nonumber \\ 
&&\ +\  \Big[{\bf v} + {\bf e}\times {\cal B}^{\prime}_{\alpha} 
- \big({\bf b}\times {\bf v}\big)\times {\cal B}^{\prime}_{\alpha}\Big]\cdot 
\partial_{R}n_\alpha,  \label{eff4}
\end{eqnarray}       
where we used abbreviated notations 
${\cal B}^{\prime}_{\alpha}
\equiv {\cal B}_{\alpha} - {\cal E}_{\alpha}\times {\bf v}$ and 
${\bf v}\equiv \partial_{k}\epsilon$~\cite{comment}. The partial  
$X$-derivative of $n_\alpha$ was taken 
with the canonical momentum $k$ fixed.  
By equating this with the
continuity equation in phase space, 
$0 = \partial_{T}n_{\alpha} + (\partial_{T}k)\cdot\partial_{k} n_{\alpha}
 + (\partial_{T}R)\cdot\partial_{R} n_{\alpha}$, 
we obtain the quasi-particle equation of motion, Eq.~(\ref{EOMquasi}), 
{\it to the accuracy of the 2nd order gradient expansion}.   
Furthermore, extracting the conserved current from the continuity
equation in zero external field (${\bf b}=0$), one finds 
${\bf j}_{\alpha} =\int_k [{\bf v} + {\bf e}\times {\cal 
  B}^{\prime}_{\alpha}]n_{\alpha}$.  The usual solution of the Boltzmann equation in linear
response therefore gives immediately the anomalous Hall conductivity in
Eq.~(\ref{eq:ahe}). 
 
\acknowledgments

Authors are pleased to acknowledge V. M. Galitski, A. A. Burkov, 
K-i. Imura and Y. Tserkovnyak.  
This work was supported by NSF Grant DMR04-57440 and 
the Packard Foundation. R.S. is supported by 
JSPS 
as a Postdoctoral Fellow.



\begin{thebibliography}{99}
\bibitem{comment2} As a simplestic example 
having finite a.e.f., one may  
consider a lattice-regularized Rashba model 
under applied magnetic field, in which 
electrons are coupled with the Einstein phonon 
having a finite mass $\omega_0$. Then, the  
2nd order perturbation with respect to this  
coupling gives rise to a  
collisional self-energy, 
with an exponentially small linf-time term $\hat{\Gamma}$ at 
low-$T$ and $|\omega-\mu| \ll \omega_0$. 
Therein, the non-zero 
$\frac{\partial \hat{\sigma}}{\partial \omega}_{|\omega=\mu}$ 
in fact brings about a finite a.e.f., which is estimated 
to be the order of $10^{0}\sim 10^{-1}\times a_{\rm lattice} \cdot t$ for 
reasonable parameters ($t$ and $a_{\rm lattice}$ 
are kinetic energy and lattice constant).
\end{thebibliography}
\end{document}